\documentclass[conference]{IEEEtran}
\IEEEoverridecommandlockouts
\usepackage{cite}
\usepackage{subfigure}
\usepackage{amsmath,amssymb,amsfonts}
\usepackage{algorithmic}
\usepackage{graphicx}
\usepackage{textcomp}
\usepackage{threeparttable}
\usepackage{xcolor}
\def\BibTeX{{\rm B\kern-.05em{\sc i\kern-.025em b}\kern-.08em
    T\kern-.1667em\lower.7ex\hbox{E}\kern-.125emX}}
\begin{document}

\title{Multi-Beam Multi-Stream
Communications for 5G and Beyond Mobile User Equipment and UAV Proof of Concept Designs
\\
}

\author{\IEEEauthorblockN{ Yiming Huo\textsuperscript{1}, Franklin Lu\textsuperscript{2}, Felix Wu\textsuperscript{3}, and Xiaodai Dong\textsuperscript{1}}

\IEEEauthorblockA{
\textsuperscript{1} Department of Electrical and Computer Engineering, University of Victoria, Canada}
\ Email: \text{ymhuo@uvic.ca, xdong@ece.uvic.ca}

\IEEEauthorblockA{
\textsuperscript{2} St. Michaels University School, Canada (Now with University of Toronto, Canada)}
\ Email: \text{franklin.lu@smus.ca, franklin.lu78@gmail.com}

\IEEEauthorblockA{
\textsuperscript{3} Department of Mechanical Engineering, University of British Columbia, Canada}
\ Email: \text{ fwfelixwu@alumni.ubc.ca}
}

\maketitle

\begin{abstract}
Millimeter-wave (mmWave), massive multiple-input multiple-output (MIMO), are expected to play a crucial role for 5G and beyond cellular and next-generation wireless local area network (WLAN) communications. Moreover, unmanned aerial vehicles (UAVs) are also considered as an important component of next-generation networks. In this paper, we propose and present a mmWave distributed phased-arrays (DPA) architecture and proof-of-concept (PoC) designs for user equipment (UE) and unmanned aerial vehicles (UAVs) which will be used in 5G/Beyond 5G wireless communication networks. Through enabling a multi-stream multi-beam communication mode, the UE PoC achieves a peak downlink speed of more than 4 Gbps with optimized thermal distribution performance. Furthermore, based on the DPA topology, the UAV aerial base station (ABS) prototype is designed and demonstrates for the first time an aggregated peak downlink data rate of 2.2 Gbps in the real-world field tests supporting multi-user (MU) application scenarios.     
\end{abstract}


\section{Introduction}
The fifth generation (5G) global deployment has been accelerated by joint efforts from both academia and industry, with millimeter-wave (mmWave) communication, massive MIMO and beamforming techniques becoming immediate reality. In particular, various promising 5G hardware components, circuits and systems have been proposed and demonstrated, from system level to device level \cite{Sadhu:5G}-\cite{Huo:5G}. On the other hand, recent years have also witnessed unmanned aerial vehicles (UAVs)/drone technologies advancing significantly, which has made them more affordable and accessible to civilian and commercial applications \cite{Zeng:UAV}-\cite{Wang:5G}. This phenomenal change gradually alters the conventional design and deployment of wireless communication infrastructures that have been dominated by the terrestrial ones since the beginning. The third Generation Partnership Project (3GPP) initialized a study item on the enhanced long-term evolution (LTE) support for UAV \cite{LTE} in March 2017, with LTE UAV field test results concluded in 3GPP Release 15. Both academia and industry have verified promising prototype designs with field tests, and UAV enabled cellular networks designs mainly in sub-6 GHz LTE bands. \textcolor{black}{However, there are few published works about system-level integration and implementation related to 5G new radio (NR) high bands (24 GHz and above) for the user equipment (UE) and UAV applications.} In this paper, we propose for the first time a mmWave distributed phased arrays (DPA) architecture based user equipment proof-of-concept (PoC) design and verification, and a mmWave DPA enabled UAV aerial base station (ABS) prototype working for multi-user (MU) application scenarios. 

\begin{figure}[!t]
\begin{center}
\includegraphics[width=3.4in]{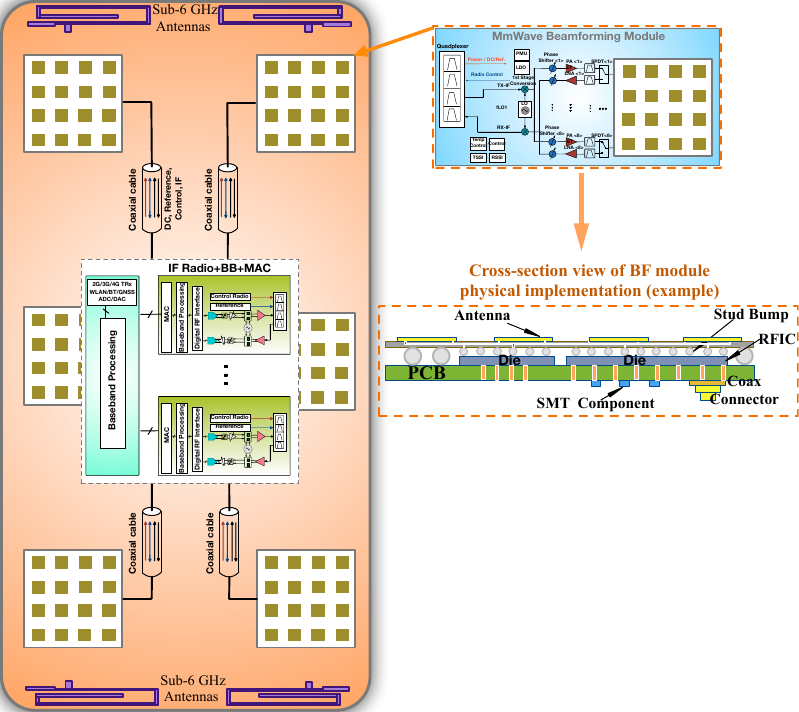}
\caption{MmWave distributed phased arrays architecture in next generation mobile devices design.}\label{FIG1}
\end{center}
\end{figure}

\section{5G USER EQUIPMENT SPECIFICATION AND DESIGN }
As illustrated in Fig.~\ref{FIG1}, the distributed phased arrays architecture proposed in \cite{Huo:5G} can mitigate several major technical challenges for future 5G hardware engineering particularly at the UE end, such as human blockage, self-heating issues, co-existence and interference cancellation (of different wireless standards) requirement. Moreover, such DPA architecture enables other advanced features, e.g., high spatial multiplexing gain, slim form factor, high reconfigurability and design freedom. Moreover, from wireless hardware design perspective, adopting split-IF architecture for mmWave transceiver designs \cite{Boers:5G} is critical to facilitate wider bandwidths, solid thermal performance and more design flexibility. 

Unlike the 5G mmWave massive MIMO base station design in which digital beamforming \cite{Yang:5G}, or hybrid beamforming \cite{Mondal:5G} with full-complexity can be afforded to use, hybrid beamforming with reduced-complexity such as DPA-MIMO is more applicable and cost-effective for the wireless UE hardware design. Moreover, the proposed DPA architecture can enable the reuse and multiplexing of both sub-6 GHz and mmWave front end modules for various application scenarios including greater carrier aggregation among 5G low bands and high bands\cite{Huo:Co-design}.  
\begin{figure}
\centering
\begin{tabular}[b]{c}
\subfigure []{\includegraphics[scale=1.0]{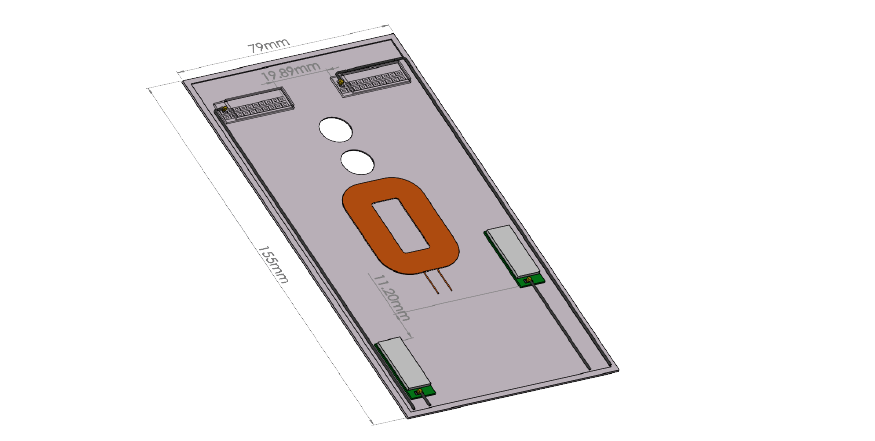}} \hspace*{0.2cm}\label{fig:FIG2A}\\
\subfigure []{\includegraphics[scale=0.7]{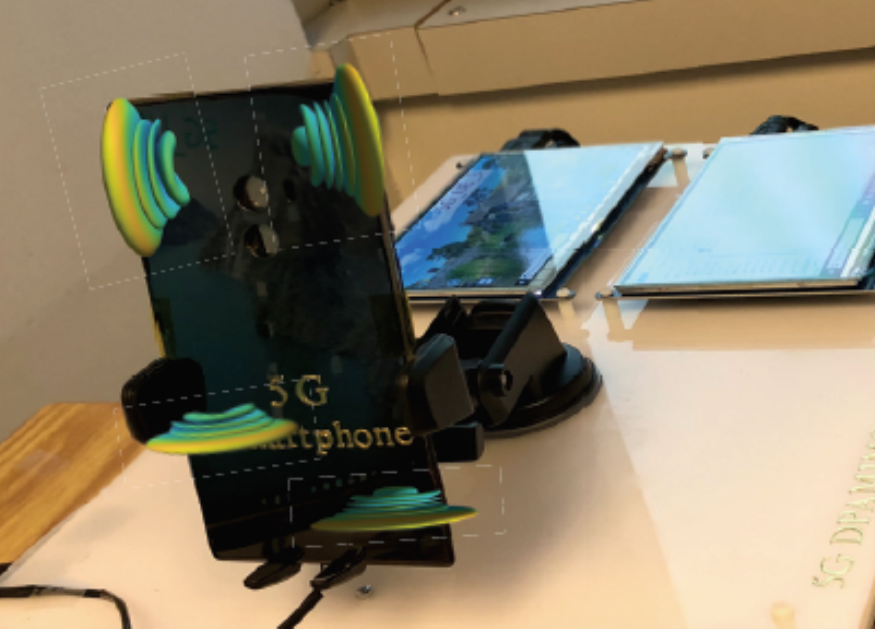}} \hspace*{0.2cm}\label{fig:FIG2B}
\end{tabular}

\caption{(a) 5G mmWave distributed phased arrays, wireless charging module, and product co-design, and (b) 5G mmWave DPA enabled PoC system implementation (with radiation pattern visualized).}\label{fig:POC}
\end{figure}

\begin{figure}
\centering
\begin{tabular}[b]{c}
\subfigure []{\includegraphics[scale=1.1]{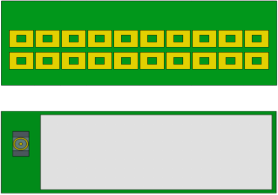}} \hspace*{ 0.2cm}\label{fig:FIG3A}
\subfigure [] {\includegraphics[scale=1.0]{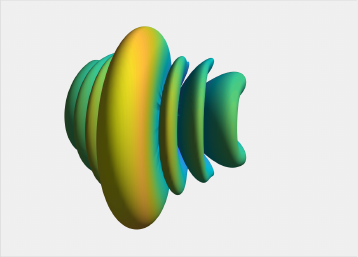}} \hspace*{ 0.2cm}\label{fig:FIG3B}\\
\end{tabular}

\subfigure [] {\includegraphics[scale=1.2]{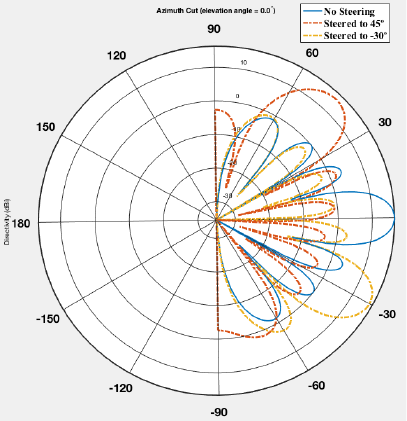}} \hspace*{ 0.2cm}\label{fig:FIG3C}\\
\subfigure [] {\includegraphics[scale=1.2]{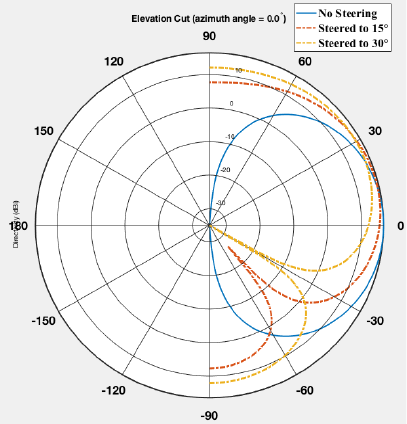}} \hspace*{ 0.2cm}\label{fig:FIG3D}
\caption{Phased array characterization (a) top and bottom view of beam-forming module, (b) 3D radiation pattern, (c) directivity of Azimuth cut (no steering, steered to 45$^{\circ}$ and -30$^{\circ}$), and (d) directivity of Elevation cut (no steering, steered to 15$^{\circ}$ and 30$^{\circ}$).}\label{fig:BFM}
\end{figure}

In our 5G UE PoC system design, a 4-layer DPA architecture is proposed, as shown in Fig.~\ref{fig:POC}. Here, four antenna beams generated by four off-the-shelf beamforming modules (BFM) respectively enable a maximum of four independent spatial data streams configured simultaneously. Furthermore, a split-IF structure is adopted to realize two-stage frequency conversion where a total of four mmWave beamforming modules working at unlicensed WiGig bands are utilized to do the first-stage frequency conversion. The second-stage frequency conversion (down to or from analog basedband), mixed-signal processing, baseband signal processing, MAC layer processing, etc., are integrated and implemented on the single system-on-chip (SoC) chipset of the WiGig adaptor which is connected to the system CPU and SDRAM through M.2 type 2230 PCI Express bus. 

To overcome the significantly large path loss and limited single antenna gain \cite{Huo:Antenna} at WiGig bands, beamforming technology needs to be used. As illustrated in Fig.~\ref{fig:BFM}(a), each WiGig beamforming module is based on multi-layer PCB design and consisted of a subarray of 2$\times$10 antenna elements (with 2$\times$1 dummy antenna elements on both sides) on the top layer, while SoC die, discrete multiplexer and I-PEX micro RF connector are placed on the bottom layer. It has a slim form factor of 25mm$\times$9mm$\times$2mm (W$\times$L$\times$D), with a weight of only 1 gram. The IF-Radio signal, reference \& control signals, DC signal (power supply), etc., are carried through the coaxial cable and I-PEX connectors. An effective isotropic radiated power (EIRP) is measured around 21 dBm for channel 1. The 3D radiation pattern, directivities of both Azimuth cut and Elevation cut are illustrated in Fig.~\ref{fig:BFM}(b)--(d), respectively. 

In order to further unveil the mmWave communication impacts on the future 5G UE product/appearance design, several mainstream materials used for manufacturing rear cases (antenna and battery back housing) of mobile handsets, namely, tempered glass, metal alloy, and ceramic, were experimented for the 5G UE PoC tests. 

\begin{table}[h]
\caption{Characterization test of candidate antenna housing materials for 5G mobile devices} \label{tab:materials}
\newcommand{\tabincell}[2]{\begin{tabular}{@{}#1@{}}#2\end{tabular}}
 \centering
 \begin{threeparttable}
 \begin{tabular}{|c|c|c|c|}\hline
        \tabincell{c}{\textbf{Item}}  &  \tabincell{c}{\textbf{Thermal}\\\textbf{Conductivity}} &  \tabincell{c}{\textbf{Hardness}\\\textbf{Mohs' scale}} &  \tabincell{c}{\textbf{Attenuation}\\\textbf{(@ 60 GHz)} } \\ 
        \hline
        
        \tabincell{c}{Glass } &\tabincell{c}{Normal} & \tabincell{c}{Good} & \tabincell{c}{Low}  \\
        \hline

        \tabincell{c}{Metal alloy } &\tabincell{c}{Best} & \tabincell{c}{Normal} & \tabincell{c}{Very high} \\
        \hline
        
        \tabincell{c}{Ceramic } &\tabincell{c}{Good} & \tabincell{c}{Good} & \tabincell{c}{Low}  \\
        \hline
   
    \end{tabular}
    
    \end{threeparttable}
\end{table}

As verified and concluded from field tests in Table I, metal alloy introduces the most significant attenuation at 60 GHz (more than 30 dB in the field tests) while ceramic rear case is, among the three, the most ideal material candidate for fabricating 5G mobile rear case by considering both the thermal conductivity and wireless performance. It is worth mentioning that thermal performance largely limits the wireless communication performance in a practical way, as the module temperature rises to a critical threshold due to the large amount of heat generation by the BFM's SoC transceivers. In recent years, aluminum nitride (AIN) ceramic draws intense attention due to its high thermal conductivity whihch is 285 W${\text{m}^{-1}·\text{K}^{-1}}$ (compared to 401 W${\text{m}^{-1}·\text{K}^{-1}}$ of copper), especially for an electrically insulating ceramic \cite{Tai:Effects}.   

Furthermore, as illustrated in Fig.~\ref{fig:POC}(a), four BFMs are arranged in a distributed way within a ceramic back housing, on top of a comprehensive consideration of both the user habit study and characteristics of the BFM. First of all, two BFMs are necessary to be accommodated alongside the top frame with an edge-to-edge spacing of 19.9 mm which is equal to around 4 times the free-space wavelength, $\lambda_0$, at 60 GHz. Such large spacing can increase antenna isolation and minimize the envelope correlation coefficient (ECC) to further improve overall data rate, which is verified in field tests. Furthermore, the ECC, $\rho_{e,12}$, between two antennas (sub-arrays) is defined in \cite{Sarkar:5G}, and can be obtained from the 3-D far-filed radiation patterns of the individual antennas from the formula \cite{Karaboikis:EEC}:  

\begin{equation} \label{ecc}
\rho_{e,12}(\omega)=\frac{|\int_{4\pi} d\Omega \; \mathbf{E}_{1}(\theta,\phi,\omega) \cdot \mathbf{E}^{*}_{2}(\theta,\phi,\omega)|^{2}}{{\int_{4\pi} d\Omega \; |\mathbf{E}_{1}(\theta,\phi,\omega)|^{2}}{\int_{4\pi} d\Omega \; |\mathbf{E}_{2}(\theta,\phi,\omega)|^{2}} },
\end{equation}
where there is assumption that the antennas (or sub-arrays) 1 and 2 are radiating in uniform multipath environment of balanced polarization\cite{Karaboikis:EEC}. The radiation patterns of antennas 1 and 2 operating at angular frequency $\omega$ are denoted by $\mathbf{E}_{1}(\theta,\phi,\omega)$ and $\mathbf{E}_{2}(\theta,\phi,\omega)$, respectively. $\int_{4\pi}$ stands for the integration with respect to the full solid angle $\Omega$. 

Second, another two BFMs are placed on each side of the bottom part of the housing to enable more diversity to cope with different application scenarios (different gestures holding the device). They are perpendicular to the top two BFMs to minimize EM coupling and cross-beam interference, with a vertical edge-to-edge spacing of 11 mm ($>$2$\lambda_0$) between each other to maintain a high antenna isolation and low ECC. 

Finally, as illustrated in Fig.~\ref{fig:POC}(b), for the 5G UE PoC design, each BFM is connected through a coaxial cable to the WiGig adaptor where the baseband/MAC layer processing is operated. Extremely fast solid-state disk (SSD) and Double Data Rate 4 Synchronous Dynamic Random-Access Memory (DDR-4 SDRAM) are used for each data stream on the UE receiver PoC. A base station (BS) PoC prototype is also configured with multiple BFMs. Field tests are conducted for two channel models of typical deployment scenarios, Indoor Hotspot (InH) Office (corridor) and Urban Micro (UMi) Street Canyon, respectively. \textcolor{black}{It is worth mentioning that, when the artificial movement within one meter range horizontally/vertically  (at 1 m/s speed with the rear housing facing the BS PoC ) is added to the test scenario to emulate the user movement, the data rate was hardly influenced. } 

\begin{figure}[!t]
\begin{center}
\includegraphics[width=3.4in]{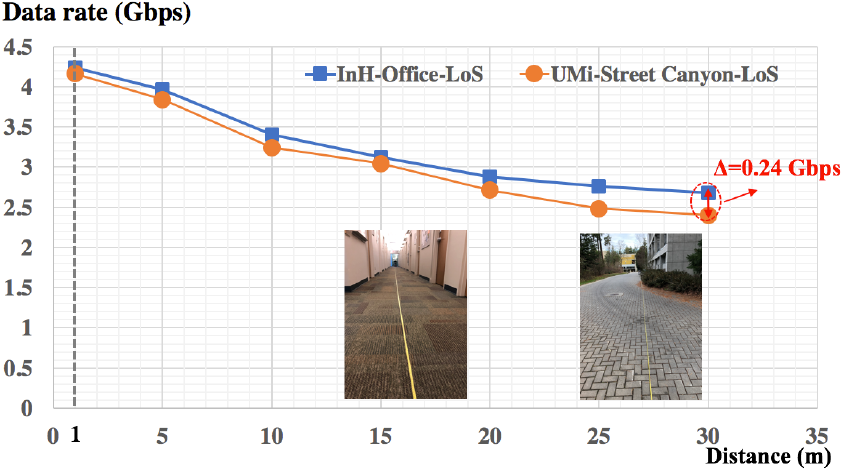}
\caption{Peak downlink data rate of 4-layer DPA versus distance for two representative channel models.}\label{FIG4}
\end{center}
\end{figure}

\begin{figure}[!t]
\begin{center}
\includegraphics[width=3.5in]{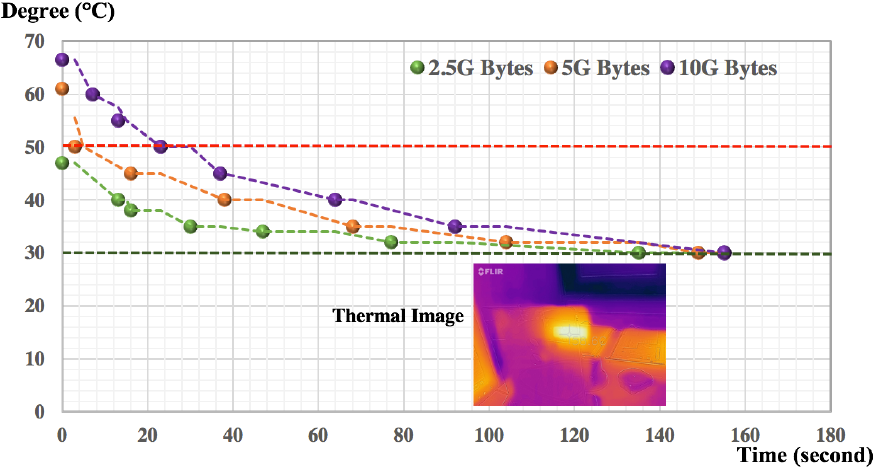}
\caption{Thermal-wireless performance of 5G ceramic smartphone antenna housing with mmWave BF module embedded.}\label{FIG5}
\end{center}
\end{figure}

As illustrated in Fig.~\ref{FIG4}, the peak downlink speed is measured at 4.23 Gbps for 4-layer DPA in an InH-Office-LoS scenario, while a slightly lower speed is measured for UMi-Street Can-yon LoS, and this difference is more pronounced for longer transmission distance. This phenomenon can be explained by close-in (CI) channel model\cite{Rappaport:5G}:    
\begin{equation} \label{pathloss}
\text{PL}^{CI}(f_\text{GHz},d)[\text{dB}]=32.4+10n\text{log}(d)+20\text{log}_{10}(f_\text{GHz})+\chi_{\sigma}^{CI},
\end{equation}
where $n$ is the path loss exponent (PLE), $d$ is the 3D distance in meters, and $\chi_{\sigma}^{CI}$ is the shadow fading (SF) term in dB. For InH-Office-LoS and UMi-Street Canyon-LoS, $n$ is 1.73 and 2.1, respectively. Moreover, the SF, $\chi_{\sigma}^{CI}$, are 3.02 dB and 3.76 dB, respectively. Consequently, InH-Office demonstrates a slightly higher speed than UMi-Street Canyon. As further shown in Fig.~\ref{FIG5}, wireless-thermal characterization test is conducted for transmitting files of various sizes, i.e. 2.5/5/10 G Bytes, in a room-temperature environment. Timing begins with the moment when the 5G UE prototype finishes receiving the files, and thermal imaging is operated right towards the surface of the rear case. 

\section{UAV SYSTEM DESIGN AND FIELD TEST }

In this section, the DPA-MIMO structure is applied to UAV communications and field test results are reported. Typically UAVs can act as aerial base stations to support ground users within the cell. However, air-to-ground (A2G) channels are different from terrestrial channels which have higher path loss exponent and heavier small-scale fading \cite{Mozaffari:UAV}. 

\begin{figure}
\centering
\begin{tabular}[b]{c}
\subfigure []{\includegraphics[scale=1.0]{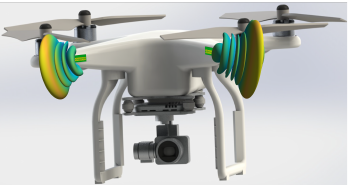}} \hspace*{0.2cm}\label{fig:FIG6A}\\
\subfigure []{\includegraphics[scale=0.45]{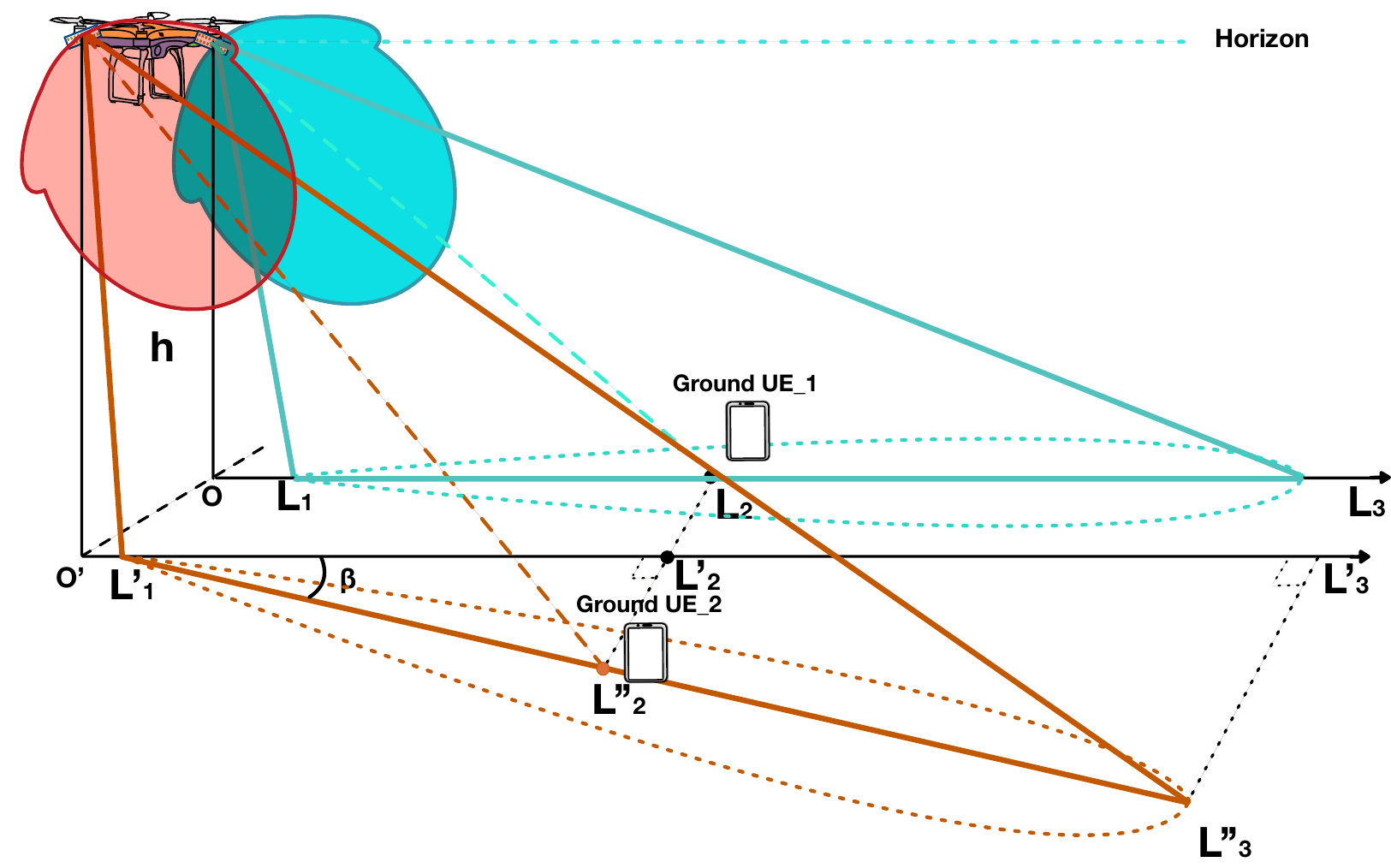}} \hspace*{0.2cm}\label{fig:FIG6B}
\end{tabular}

\caption{UAV-ABS system with DPA enabled mmWave communication and (b) 3D view of multi-user application scenario of UAV-ABS.}\label{fig:FIG6}
\end{figure}

Referring to Fig.~\ref{fig:FIG6}(a), in this work, multi-beam multi-stream mmWave communication scheme and architecture are adopted for UAV-ABS’s MU application scenario. As shown in Fig.~\ref{fig:FIG6}(b), from a 3D view, two ground UEs are separated with a distance $L_{\text{2}}$-$L^{''}_{\text{2}}$, and $\beta$ is the intersection angle formed by the projection of two BFMs’ main lobes in the elevation pattern, on the ground. Straight lines $L_{\text{3}}$-$L_{\text{1}}$ and $L^{''}_{\text{3}}$-$L^{'}_{\text{1}}$ stand for the projection of the maximum elevation gain of two BFMs on the ground, respectively. With a suitable angle $\beta$ obtained via steering the two beams, cross-beam interference can be significantly mitigated.   

\begin{figure}
\centering
\begin{tabular}[b]{c}
\subfigure []{\includegraphics[scale=1.2]{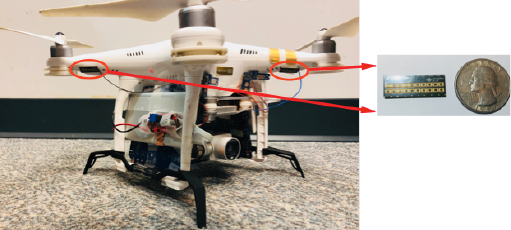}} \hspace*{0.2cm}\label{fig:FIG7A}\\
\subfigure []{\includegraphics[scale=1.1]{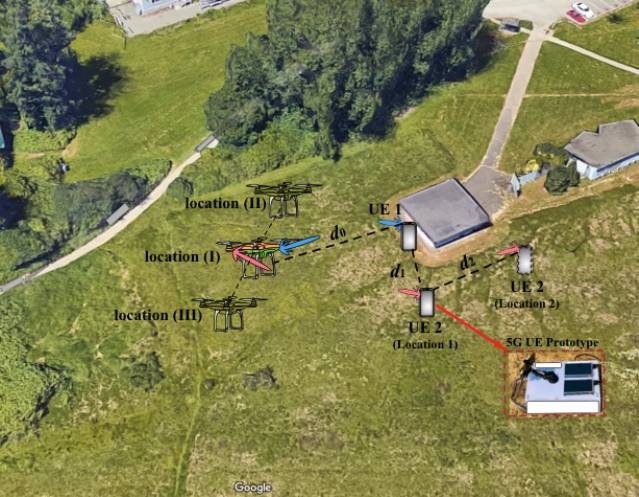}} \hspace*{0.2cm}\label{fig:FIG7B}
\end{tabular}

\caption{(a) UAV-ABS hardware implementation and (b) filed trial tests with various application scenarios.}\label{fig:FIG7}
\end{figure}

Furthermore, the system proof of concept design and implementation is subject to the UAV's extra allowable payload as well as limited onboard power supply. As illustrated in Fig.~\ref{fig:FIG7}(a), two 60 GHz BFMs are fixed onto two arms of the DJI Phantom 3. The main logic boards (MLBs) that accommodate IF radio and baseband stage circuitry are attached onto the UAV body with high-efficiency DC-DC converters and Li-ion batteries. The overall extra payload is maintained at around 1.2 pound including USB 3.0 flash drives providing test files. 

Field tests are conducted in a local area shown in the Google’s 3D map of Fig.~\ref{fig:FIG7}(b). There are two typical scenarios where the UAV-ABS provides mmWave enabled gigabits-per-second (Gbps) link service, i.e., single-user (SU) and MU mode. For SU, the UAV-ABS hovers at a specific height and enables two beams to transmit two data streams to ground UE prototype which has been implemented and presented in the previous section. With two BFMs activated on one single ground UE, a peak data rate of 2.24 Gbps is achieved. When the UAV-ABS hovers around 41 m away from the ground UE ($h$ is 35 m, and $d_\text{0}$ is 22 m), the communication link is still very stable. It is worth mentioning that, compared to terrestrial communications when the signal strength and connection become weak and unstable (e.g., UMi-Street Canyon), as the distance goes over 30 m, A2G communication demonstrates advantages, which complies with the A2G channels characteristics holding smaller path loss exponent and lighter small-scale fading. In the MU scenario tests, two ground UEs with single BFM enabled on each are deployed with separation distance, $d_1$, varied to receive individual beam ($d_1$ and $d_0$ are perpendicular, and $d_1$ is set to 6 m, and 2 m, respectively). UE 1 and UE 2 achieve an aggregated peak data rate of 2.168 Gbps (1.096 Gbps and 1.072 Gbps for UE 1 and UE 2, respectively). In another test scenario, when UE 2 is placed at Location 2 with $d_2$ ($d_0$ and $d_2$ are parallel) set to 10 m and the other parameters are the same as before, the aggregated data rate is slightly decreased to 2.136 Gbps. Therefore, wider separation of ground users can result in a larger intersection angle, $\beta$, which is beneficial to less cross beam interference. On the other hand, each phased array has a finite steering angle range, and wider separation of users may require both phased arrays to steer to steering angle bounds that lead to lower BFM gains.   

On the other hand, the quasi-stationary status of UAV in which small movement and slow horizontal drifting happens (e.g. from Location I to Location II, or Location I to Location III, as shown in Fig. 7 (b)) due to limited localization accuracy and abrupt atmospheric pressure change (when the wind speed is 6-7 m/s reported on ground), does not obviously affect the performance of the wireless communications in the field tests. This can be explained by, first, the aerodynamic system can facilitate necessary system-level robustness and reliability against environmental variance; second, the beamforming and wireless design are appropriate and assist solving the environment-induced impairments. In this case, a 10-degree or so HPBWA can tolerate a large horizontal shift compared to the distance. Moreover, the beam misalignment can be quickly fixed by beam tracking and re-alignment from the MAC-layer algorithms.

\section{CONCLUSION}
This paper presents proof-of-concept design of a distributed phased arrays architecture for next-generation mobile user equipment and UAV applications. \textcolor{black}{As compared to conventional single-stream single-beam mmWave system design, DPA architecture demonstrates several merits particularly the significant improvement in the data throughput.} As verified from measurements and field tests, the mmWave DPA enabled UE prototype can achieve a peak downlink data rate more than 4.2 Gbps with satisfactory thermal-wireless performance. Moreover, mmWave DPA architecture further presents its advantages for the multi-beam multi-stream capable UAV-ABS system design. Field tests shows a peak downlink data rate of 2.2 Gbps for SU and MU application scenarios, with UAV's immunity to environment-induced impairments such as mildly challenging weather. \textcolor{black}{The future work will focus on several aspects: first, further increasing the data throughput of the mobile user equipment under mobile and other challenging deployment scenarios; second, investigating the UAV body movement (due to the weather) induced impacts on A2A/A2G mmWave communications and beamforming design.}

\end{document}